\documentclass[aps,10pt,notitlepage,twocolumn,nofootinbib,superscriptaddress]{revtex4-2}

\usepackage{graphicx,color}
\usepackage{amsmath}
\usepackage{amssymb}
\usepackage{hyperref}
\usepackage[utf8]{inputenc}
\usepackage[english]{babel}
\usepackage{epsfig}
\usepackage{subfigure}
\usepackage{wasysym}
\usepackage{color,xcolor}
\usepackage{amsmath}
\usepackage{bm}
\usepackage{epsfig}
\usepackage{amsfonts}
\usepackage{dcolumn}
\usepackage{float}
\usepackage{booktabs}
\usepackage{relsize}

\hypersetup{colorlinks=true, linkcolor=blue, citecolor=green}

\usepackage{dcolumn}
\usepackage{bm}
\usepackage{ifpdf}
\usepackage{hyperref}
\usepackage{float}
\usepackage{bm}
\usepackage{xcolor,color,graphicx,graphics}
\usepackage[OT1]{fontenc}
\usepackage{latexsym,amssymb,amsmath,amsfonts}
\usepackage{makeidx}
\usepackage{epsfig}
\usepackage{epstopdf}
\usepackage{mathrsfs}
\hypersetup{colorlinks=true, linkcolor=blue, citecolor=green}
\usepackage{enumerate}
\usepackage{xcolor}
 \usepackage{multirow}

\begin{document}

\title{Relativistic structure of a supermassive black hole embedded\\ in the dark matter halo of NGC 4649 (M60)}

	\author{Francisco S. N. Lobo} \email{fslobo@ciencias.ulisboa.pt}
\affiliation{Institute of Astrophysics and Space Sciences, Faculty of Sciences, University of Lisbon, Building C8, Campo Grande, P-1749-016 Lisbon, Portugal}
\affiliation{Department of Physics, Faculty of Sciences, University of Lisbon, Building C8, Campo Grande, P-1749-016 Lisbon, Portugal}

     \author{Jorde A. A. Ramos}
    \email{jordefisica@gmail.com}
\affiliation{Faculty of Physics, Graduate Program in Physics, Federal University of Pará, 66075-110, Belém, Pará, Brazil}

    \author{Manuel E. Rodrigues} \email{esialg@gmail.com}
\affiliation{Faculty of Physics, Graduate Program in Physics, Federal University of Pará, 66075-110, Belém, Pará, Brazil}
\affiliation{Faculty of Exact Sciences and Technology, Federal University of Pará, Abaetetuba University Campus, 68440-000, Abaetetuba, Pará, Brazil}


\begin{abstract}

We construct a static, spherically symmetric black hole (BH) solution embedded within a dark matter (DM) halo, formulated as a non-vacuum extension of the Schwarzschild spacetime. The DM distribution is modeled via an empirical density profile calibrated to observations of the elliptical galaxy NGC 4649 (M60), incorporating Hubble Space Telescope (HST) imaging, stellar velocity dispersion data, and globular cluster dynamics. The resultant spacetime metric depends on three independent parameters: the black hole mass $M$, the asymptotic circular velocity $V_c$, and the halo scale radius $a$, and smoothly reduces to the Schwarzschild limit as $V_c \to 0$ and $a \to 0$.
We analyse the influence of the halo on key geometric and physical quantities, including the event horizon radius, photon sphere, shadow size, and curvature invariants. The Kretschmann scalar exhibits an enhanced sensitivity to halo-induced modifications, particularly in the near-horizon regime.
Thermodynamic properties of the solution are also examined. In the extremal limit, characterized by a vanishing surface gravity, the model supports a finite tangential pressure, implying a non-trivial extension of standard black hole thermodynamics. These results highlight the relevance of incorporating astrophysical environments into BH modelling and offer new avenues for testing strong-field gravity through precision observational data.

\end{abstract}
\date{\today}
\maketitle
\def\HMS{{\scriptscriptstyle{\rm HMS}}}

\section{Introduction}\label{sec1}

Black holes (BHs) have long served as natural laboratories for probing the interplay between gravity and quantum theory, one of the deepest challenges in theoretical physics. The pioneering formulation of the Schwarzschild solution in General Relativity (GR) \cite{Schwarzschild:1916uq} laid the foundation for understanding these enigmatic objects. Further advances in quantum field theory in curved spacetimes emphasized BHs as critical probes of fundamental physics. A major milestone came with Hawking’s discovery that BHs emit radiation \cite{Hawking:1975vcx}, leading to the realization that they are thermodynamic systems governed by laws analogous to classical thermodynamics \cite{Bardeen:1973gs,Wald:1984rg}. These findings laid the foundation for the field of BH thermodynamics, in which classical thermodynamic quantities such as entropy, temperature, and energy acquire profound geometric and quantum interpretations \cite{Gibbons:1977mu,Davies:1977bgr,Toussaint:1978br,York:1986it,Gibbons:1996af}.

Concurrently, astrophysical and cosmological observations over the past few decades have made it increasingly clear that the visible matter in the universe accounts for only a fraction of its total mass-energy content. The bulk is attributed to a non-luminous, non-baryonic component known as dark matter (DM) \cite{Bertone:2004pz,Planck:2018vyg,Freese:2008cz,deSwart:2017heh,Wechsler:2018pic}. Despite the success of the $\Lambda$CDM model in explaining large-scale structure formation, the true nature of DM remains elusive \cite{Arbey:2021gdg,Cebrian:2022brv,Misiaszek:2023sxe}. Among the diverse candidates proposed, ranging from WIMPs to axions and other exotic particles \cite{Steffen:2008qp,Zurek:2024qfm,Feng:2010gw}, it is now well recognized that DM must also play a central role in the dynamics and evolution of galaxies and their central supermassive black holes (SMBHs) \cite{EventHorizonTelescope:2022wkp,EventHorizonTelescope:2019dse,Valluri:2002xs,Marasco:2021pkl}.

This recognition has motivated a growing body of research into how BHs interact with surrounding DM halos. Unlike idealized models of isolated BHs in asymptotically flat spacetimes, astrophysical BHs reside in complex galactic environments shaped by DM \cite{Cardoso:2021wlq, Heydari-Fard:2024wgu, Shen:2024qbb}. Recent theoretical and numerical studies show that such environments can significantly influence BH behaviour. In particular, DM may affect accretion dynamics, perturb the horizon geometry, and modify key thermodynamic quantities \cite{Xu:2018wow, Liu:2022lrg, Gondolo:1999ef, Sadeghian:2013laa}. These interactions may also impact BH growth and feedback over cosmological timescales. Collectively, these findings highlight the need to account for DM in any realistic model of BH physics, especially when interpreting observations or testing gravity in strong-field regimes.

Moreover, DM density profiles, including the widely used Navarro-Frenk-White (NFW) model \cite{Navarro:1996gj}, along with its various alternatives \cite{deBlok:2009sp, Tran:2024hry, Hernquist:1990be}, have played a vital role in modelling the gravitational potential on galactic scales. These profiles provide crucial input for understanding the structure and dynamics of galaxies. When embedded into relativistic spacetime geometries, such profiles enable the construction of BH metrics that exhibit significant departures from the standard Schwarzschild or Kerr solutions \cite{Shen:2009my, Vagnozzi:2022moj, Ma:2024oqe}. These generalized configurations offer a powerful framework for studying BHs not only as high-energy astrophysical engines, but also as precision probes of DM distributions and their underlying physical properties \cite{Padilla:2020sjy, El-Zant:2020god, Bansal:2022enx}.

In this work, we explore BH solutions embedded within DM halos, with a particular focus on the SMBH residing in the elliptical galaxy NGC 4649 (M60). Our analysis incorporates observational data from the Hubble Space Telescope, complemented by stellar kinematic measurements and globular cluster dynamics. NGC 4649 is a giant elliptical galaxy characterized by low surface brightness and situated in a substructure to the east of the main Virgo cluster concentration \cite{Forbes:2004kp}. It has been the subject of extensive studies aimed at reconstructing its total mass profile, using both X-ray observations \cite{Humphrey:2006rv, Gastaldello:2006kw, Humphrey:2008dz} and globular cluster system analyses \cite{Bridges:2006hu}. These comprehensive datasets provide a valuable framework for jointly constraining the properties of the central BH and the surrounding DM distribution.

A key observational feature considered in our study is the BH shadow, a silhouette formed by the gravitational lensing of light in the strong-field region near the event horizon \cite{Vagnozzi:2022moj}. The recent imaging of the shadow of Sagittarius A* by the Event Horizon Telescope (EHT) collaboration has opened new avenues for probing the near-horizon structure of BHs and testing GR in extreme regimes. Theoretical models explaining shadow morphology have encompassed a broad range of scenarios, including regular BHs, metrics inspired by string theory, and modifications to GR. Building on these developments, we examine BH solutions embedded in DM environments with a logarithmic
profile by adopting the density profile $\rho_{\mathrm{DM}}$ proposed in \cite{Shen:2009my}, and given by
\begin{equation}
	\rho_{\mathrm{DM}}(r) = \frac{V_c^2}{4\pi G}\,\frac{3a^2 + r^2}{(a^2 + r^2)^2} \label{densidade} \ ,
\end{equation}
where $a$ denotes the characteristic core radius of the halo, and $V_c$ is the critical tangential velocity associated with the DM distribution.
The spacetime geometry is constructed following the methods outlined in \cite{Cardoso:2021wlq}, while observational constraints on the shadow shape are incorporated using results from \cite{Vagnozzi:2022moj}.

In our analysis, we chose not to adopt the traditional NFW profile. Although this model provides a realistic fit to the kinematics of globular clusters (GCs), it lacks the necessary central concentration to account for mass dominance in the vicinity of the BH \cite{Shen:2009my, Gebhardt2009Apj}. While the NFW profile does exhibit a rising density toward the center, this increase is insufficient to accurately represent the central mass distribution. Alternatively, the density profile $\rho_{\rm DM}$, given by Eq.~(\ref{densidade}), not only provides a statistically robust fit to the data from the M60 galaxy \cite{Shen:2009my}, supporting analyses involving the parameters $a$ and $V_c$, but also exhibits a stable, non-divergent behavior as $r \to 0$. This indicates the presence of a central \textit{core} rather than a \textit{cusp}, a feature that aligns more closely with observational data. The regular behavior at the center is a desirable property in density models, as it enhances consistency with empirical observations \cite{Tran:2024hry}.

This paper is organized as follows: In Section~\ref{sec2}, we present the generalized spacetime metric, the adopted DM density profile, and the Kretschmann scalar, and introduce the BH solutions incorporating DM halos. Section~\ref{sec:propriedades} offers numerical estimates for the core radius, velocity profiles, and the BH mass, followed by a detailed analysis of the properties of the solutions. We conclude with a summary and outlook in Section~\ref{sec:concl}.

Throuhgout this work, we consider the metric signature $(-,+,+,+)$ and geometrized units ($G = c = 1$).

\section{Black Hole Solution embedded in a Dark Matter Halo}\label{sec2}

\subsection{Spacetime metric}

In the context of static and spherically symmetric solutions, it is generally appropriate to consider that $g_{tt} \neq -1/{g_{rr}}$, as the equality arises only in specific cases \cite{Dinverno:1990}. Under this more general assumption, the line element can be expressed in the following form:
\begin{equation}
	ds^2 = -F(r)\,dt^2 + \frac{1}{G(r)}\,dr^2 + r^2\,d\Omega^2 \label{metrica} \ ,
\end{equation}
where the angular surface element in spherical coordinates is given by
$d\Omega^2 = d\theta^2 + \sin^2{\theta}\,d\phi^2$. In the general case, $F(r)$ and $G(r)$ are functions of the radial coordinate. The Schwarzschild metric represents a specific case in which the metric functions satisfy $F(r) = G(r) = 1 - 2m_{\rm BH}/r$, where $m_{\rm BH}$ denotes the constant mass of the BH \cite{Schwarzschild:1916uq}. By relaxing the condition $F(r) = G(r)$, one obtains generalized extensions of the Schwarzschild solution. In the following sections, we will construct a solution where these functions differ.

In metric solutions describing static and asymptotically flat spacetimes, event horizons correspond to non-singular physical surfaces \cite{Morris:1988}. These are characterized by the vanishing of the temporal component of the metric\footnote{In an asymptotically flat spacetime, the surface defined by the vanishing of the normalized square of the temporal Killing vector $\xi^{\mu}$, that is, $\xi^{\mu} \xi_{\mu} = g_{tt} = 0$, constitutes a null hypersurface. This surface functions as an event horizon, beyond which no future-directed causal curve, tangent to null geodesics confined within that region, can escape \cite{Morris:1988}.}, i.e.,
\begin{eqnarray}
    g_{tt}\to0 \ .
\end{eqnarray}
This occurs because the Killing vector becomes null at the horizon, aligning with the direction of null geodesics and remaining tangent to them, thereby confined to the event horizon surface itself \cite{Vishveshwara:1968}.

The approach adopted in this work builds upon the method developed from the initial framework presented in \cite{Cardoso:2021wlq}, which allows for the construction of static, spherically symmetric BH solutions embedded within external DM fields. In the absence of a surrounding DM halo, the resulting spacetime naturally reduces to an extension of the standard Schwarzschild BH solution. This ensures consistency with classical GR in the appropriate limit. 

To model the influence of DM, we assume a specific density profile for the halo, as given by Eq.~\eqref{densidade}, which captures the essential features of realistic galactic DM distributions and enables a systematic analysis of its effects on the BH geometry.

\subsection{Kretschmann scalar}

Several properties of the BH solutions can be analyzed by the Kretschmann scalar $K = R_{\mu\nu\lambda\phi}R^{\mu\nu\lambda\phi}$, where $R_{\mu\nu\lambda\phi}$ denotes the Riemann tensor \cite{Lobo:2020ffi}. For the metric~(\ref{metrica}), the scalar $K$ has the general form,
\begin{eqnarray}
	K &=& \frac{1}{4 r^4 f(r)^4}
	\Big[r^4 g(r)^2 f'(r)^4-2 r^4 f(r)
	g(r) f'(r)^2 \times
	\nonumber
	\\
	&&\hspace{-1cm}\times \Big(2 g(r)
	f''(r)
	+f'(r)
	g'(r)\Big)+f(r)^2 \Big(4 r^4 g(r)^2
	f''(r)^2 \nonumber
	\\
	&& \hspace{-1cm} +r^2 f'(r)^2 \big(r^2
	g'(r)^2
	+8 g(r)^2\big)
	+4 r^4 g(r)
	f'(r) f''(r) g'(r)\Big)
	\nonumber
	\\
	&&+8 f(r)^4 \left(r^2 g'(r)^2+2
	(g(r)-1)^2\right)\Big] \,.
	 \label{K_escalar_00}
\end{eqnarray}

\subsection{Specific Solution}

Throughout this work, we adopt the approach proposed in~\cite{Cardoso:2021wlq}, which leads to the following line element:
\begin{equation}
	ds^2 = -F(r)\,dt^2 + \frac{dr^2}{1 - \frac{2M(r)}{r}} + r^2 d\Omega^2 \ ,\label{metrica_semidef}
\end{equation}
where the metric represents a static, spherically symmetric spacetime of the Schwarzschild type, modified by the presence of a DM halo composed of massive particles in stable, concentric orbits around the BH.

The spherically symmetric spacetime described in Eq.~(\ref{metrica}) can also be expressed in the form given by Eq.~(\ref{metrica_semidef}), where the functions $F(r)$ and $M(r)$ are introduced explicitly. By setting $F(r) = 1 - 2m_{\rm BH}/r$ and $M(r) = m_{\rm BH}$, Eq.~(\ref{metrica_semidef}) reduces to the Schwarzschild solution. In what follows, we derive a Schwarzschild-type solution based on the density profile given in Eq.~(\ref{densidade}), which incorporates the presence of a dark matter (DM) halo. For specific parameter choices, this solution reduces to the standard Schwarzschild form.

In this framework, the gravitational influence of the BH induces a pressure within the halo that is purely angular and isotropic, resulting in a pressure function that depends only on the radial coordinate. These features are encapsulated in the anisotropic energy-momentum tensor employed in constructing the solution:
\begin{equation}
    G^{\mu}_{ \ \nu}=8\pi\left(T^{\rm (DM)}\right)^{\mu}_{ \ \nu} \,,
\end{equation}
where $\left(T^{\rm (DM)}\right)^{\mu}_{ \ \nu}={\rm diag}(-\rho_{\rm DM}, 0,P,P)$, and $P$ is the tangential pressure. The relationships involving the density profile $\rho_{\rm DM}=\rho_{\rm DM}(r)$, the mass function $M(r)$, the metric coefficient $F(r)$ and the tangential pressure function $P=P(r)$ are found from the components of the energy-momentum tensor. Thus, setting $\left(T^{\rm (DM)}\right)^{t}_{ \ t}=-\rho_{\rm DM}(r)$, we have
\begin{equation}
M'(r)=4\pi r^2\rho_{\rm DM}(r)\label{massa_met3} \,.
\end{equation}
For $\left(T^{\rm (DM)}\right)^{r}_{ \ r}=0$, where we consider the radial pressure to be zero due to the absence of radial motion of the matter constituting the halo, we have
\begin{equation}
\frac{F'(r)}{F(r)}=\frac{2M(r)}{r[r-2M(r)]} \label{eqf_met03} \,.
\end{equation}
Finally, for $\left(T^{\rm (DM)}\right)^{\theta}_{ \ \theta}=\left(T^{\rm (DM)}\right)^{\phi}_{ \ \phi}=P(r)$, where using Eqs.~(\ref{massa_met3}) and~(\ref{eqf_met03}), we obtain
\begin{equation}
P(r)=\frac{M(r)\rho_{\rm DM}(r)}{2[r-2M(r)]} .\label{pressao_03_DEF}
\end{equation}
By solving the coupled field equations~(\ref{massa_met3})--(\ref{pressao_03_DEF}), we obtain the functions $M(r)$, $F(r)$ and $P(r)$.

Substituting the energy density~(\ref{densidade}) in Eq.~(\ref{massa_met3}), leads to the mass function
\begin{equation}
M(r)=m_{\rm BH} + \frac{r^3V_c^2}{a^2+r^2}\label{m(r)_met03} \ ,
\end{equation}
where the mass of the BH, $m_{\rm BH}$, results from an integration constant $C_1$, i.e. $C_1=m_{\rm BH}$. Thus, if $V_c=0 \to M(r)=m_{\rm BH}$.
If we now insert the mass function~(\ref{m(r)_met03}) into Eq.~(\ref{eqf_met03}), we obtain $F(r)$
\begin{equation}
    F(r)=\frac{C_2(r-\lambda)^{\xi}}{r} \ , 
\end{equation}
where 
\begin{equation}
    \xi=\frac{(a^2+\lambda^2)}{a^2-4m_{\rm BH}\lambda+3\lambda^2(1-2V_c^2)}
\end{equation}
and the constant of integration is $C_2=1$. 

The term $\lambda$ is defined by solving the equation
\begin{equation}
2a^2m_{\rm BH}-a^2\lambda+2m_{\rm BH}\lambda^2-\lambda^3(1-2V_c^2)=0 
\end{equation}
which leads to an expression with $\lambda\in\mathbb{R}$, given by
\begin{equation}
\lambda=\frac{A_{\lambda}}{3 \left(2
   V_c^2-1\right)}+\frac{a^2}{A_\lambda}+\frac{4
   m_{\rm BH}^2}{3 \left(2
   V_c^2-1\right)A_{\lambda}}-\frac{4 m_{\rm BH}}{6 \left(2
   V_c^2-1\right)} \ ,
\end{equation}
where
\begin{eqnarray}    
	A_{\lambda}^3 &=& \Bigr\{27\left(a-2 a V_c^2\right)^2 \Bigr[a^4\left(1-2 V_c^2\right) 
	\nonumber \\
	&&+ 4 a^2 m_{\rm BH}^2\Bigr(27 V_c^4
	-18 V_c^2+2\Bigr) +16 m_{\rm BH}^4\Bigr]\Bigr\}^{1/2} 
	\nonumber \\
	&&-8 m_{\rm BH}^3-18 a^2 m_{\rm BH} \Bigr(6V_c^4
	-5V_c^2+1\Bigr) \ .
\end{eqnarray}
For this solution, the line element is therefore defined as
\begin{equation}
ds^2=-\frac{(r-\lambda)^{\xi}}{r}dt^2+\frac{dr^2}{1-\frac{2}{r}\left(m_{\rm BH} + \frac{r^3V_c^2}{a^2+r^2}\right)}+r^2d\Omega^2\label{ds_sol03} \,.
\end{equation}
The Schwarzschild solution is recovered in the limiting case $(V_c, a) = (0, 0)$, which corresponds to $\lambda = 2m_{\rm BH}$ and $\xi = 1$.

\section{Geometric and Thermodynamic Properties}\label{sec:propriedades}

In this section, we adopt the values for the mass $M$, critical velocity $V_c$, and critical radius $a$ as defined in~\cite{Shen:2009my}. These parameters define the intervals listed in Table~\ref{tab:parametros}, which include both SI units and geometrized units for consistency and completeness.

We distinguish between two primary datasets derived from that study. The first dataset, hereafter referred to as \textbf{Data I}, is based on X-ray observational data. The second dataset, denoted as \textbf{Data II}, is derived from the dynamical mass profile modeling. The parameter values corresponding to each dataset are as follows:
\begin{itemize}
	\item \textbf{Data I}: $V_{c}^{(I)} = 13.68 \times 10^{-4}$, $a^{(I)} = 30.86 \times 10^{19}\,\text{m}$, and $M^{(I)} = 5.17 \times 10^{12}\,\text{m}$;
	\item \textbf{Data II}: $V_{c}^{(II)} = 18.35 \times 10^{-4}$, $a^{(II)} = 46.29 \times 10^{19}\,\text{m}$, and $M^{(II)} = 6.65 \times 10^{12}\,\text{m}$.
\end{itemize}
These values serve as input for our numerical and analytical investigations of BHs embedded in DM halos.

\begin{table}[h]
    \centering
\renewcommand{\arraystretch}{1.2}
    \begin{tabular}{lcc}
        \toprule
        \textbf{Parameter} & \textbf{Minimum} & \textbf{Maximum} \\
        \midrule
        $V_c$[km/s] (IS) & $300$ & $1100$ \\
        $V_c$ (Geom.) & $10 \times 10^{-4}$ & $36.7 \times 10^{-4}$ \\
        $a$[kpc] (IS) & $5$ & $90$ \\
        $a$[m] (Geom.) & $15.43 \times 10^{19}$ & $277.74 \times 10^{19}$ \\
        $M$[$M_{\odot}$] (IS) & $2.5 \times 10^{9}$ & $6 \times 10^{9} $ \\
        $M$[m] (Geom.) & $3.69 \times 10^{12}$ & $8.86 \times 10^{12}$ \\
        \bottomrule
    \end{tabular}
    \caption{Parameters $V_c$, $a$ and $M$ in different units with minimum and maximum values. The IS values were taken from~\cite{Shen:2009my}.}
    \label{tab:parametros}
\end{table}

As reported in~\cite{Shen:2009my}, the mass of the BH in the absence of a surrounding DM halo is given by $M_0 = 4.3 \times 10^{9} M_{\odot}$, which corresponds to $M_0 = 6.35 \times 10^{12}\,$m in geometrized units.

An additional parameter relevant for analyzing the properties of the solutions is the distance from the observer (on Earth) to the galaxy M60. This observational baseline is given by $r_O = 15.7 \times 10^{3}\,$kpc, which is equivalent to $r_O = 48.45 \times 10^{22}\,$m.

\subsection{Event Horizon}

To obtain the event horizon radius $r_h$, we solve $F(r)=0$ for
\begin{equation}
 F(r)=\frac{(r-\lambda)^{\xi}}{r} \ ,\label{eq_f_sol_03}
\end{equation}
which yields
\begin{equation}
    r_h=\lambda . \label{rh3}
\end{equation}
In accordance with the intervals of $M, V_c$ and $a$, we have $\xi \geq 1$, where $r_h = 2m_{BH} \to \xi = 1$. These conditions for $\xi$ imply
\begin{equation}
0<V_c\leq\frac{2}{3}\label{Vbaixasol03} \ ,
\end{equation}
and 
\begin{equation}
a>\frac{2m_{BH}}{\sqrt{3(1-2V_c)}}\ .\label{a_regiao_sol3}
\end{equation}
As shown in Eq.~(\ref{Vbaixasol03}), this model predicts a low critical tangential velocity for the orbital radius of the halo, consistent with the observational data reported in \cite{Shen:2009my}. This indicates that higher velocities may lead to dynamical instability. Furthermore, Eq.~(\ref{a_regiao_sol3}) demonstrates that as $V_c$ increases, the range of admissible values for the radius $a$ decreases, though this reduction becomes progressively more significant.

\begin{figure}[tb!]
\includegraphics[scale=0.235]{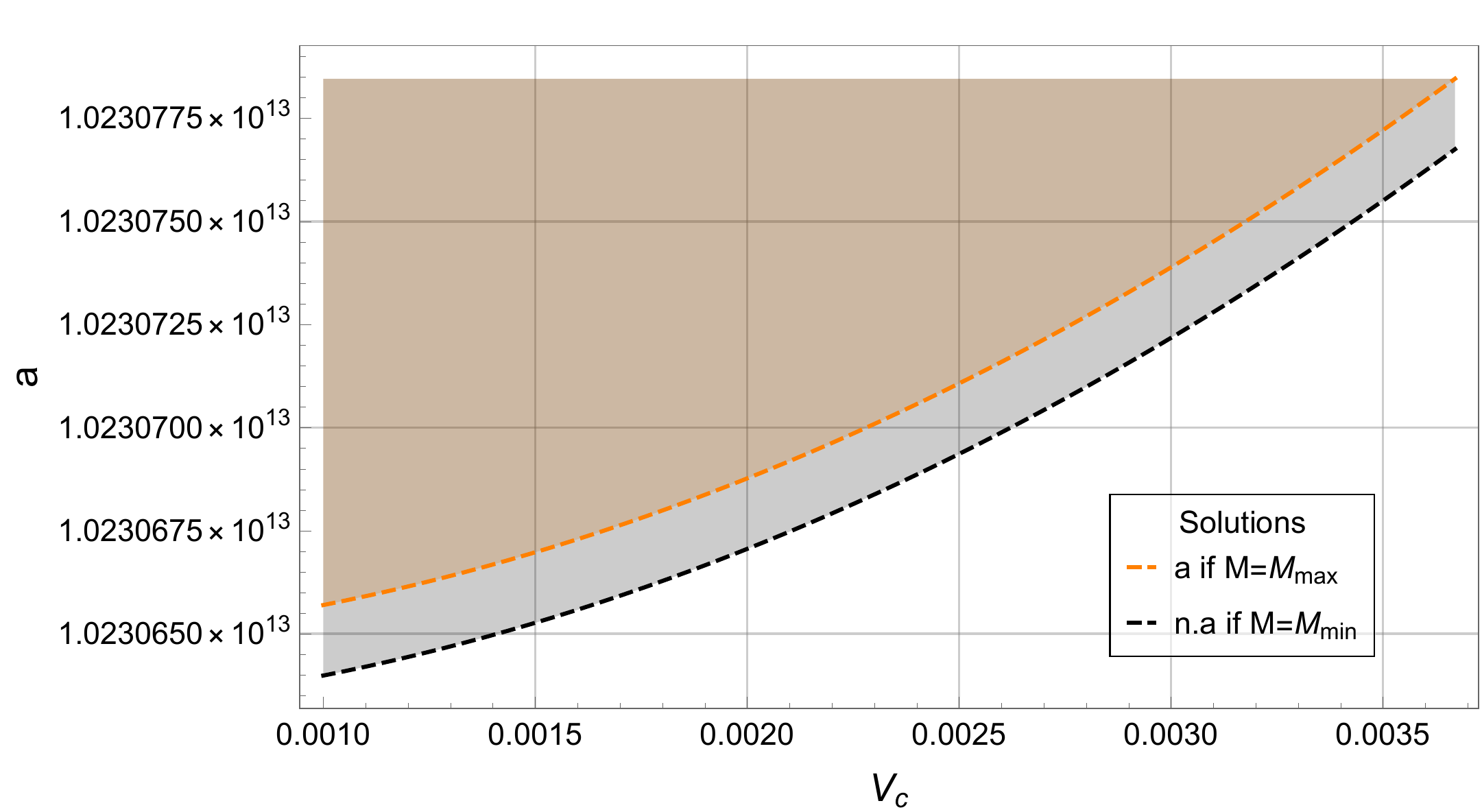}
\caption{Graphical representation of $a=a(V_c)$, from Eq.~(\ref{a_regiao_sol3}), due to the boundary condition $\xi>1$. The constant $n=M_{max}/M_{min}$, where $M_{max}=8.86\times10^{12}$m and $M_{min}=3.69\times10^{12}$m, come from the intervals originally presented in Table \ref{tab:parametros}.} 
\label{fig:a_vs_Vc_model03}
\end{figure}

Figure~\ref{fig:a_vs_Vc_model03} illustrates that the constraints on the parameter $a$ become more restrictive for more massive black holes (BHs), meaning that the permissible range for $a$ narrows as the BH mass increases. This behavior is not only consistent with the observational data presented in \cite{Shen:2009my}, but is also physically expected, as more massive BHs tend to influence matter at larger characteristic radii.
For Data I and II we have: $a_{I}>5.96\times10^{12}$, and $\xi_{I}-1=1.26\times10^{-20}$; and $a_{II}>7.67\times10^{12}$, and $\xi_{II}-1=1.67\times10^{-20}$.

\begin{figure}[tb!]
\includegraphics[scale=0.30]{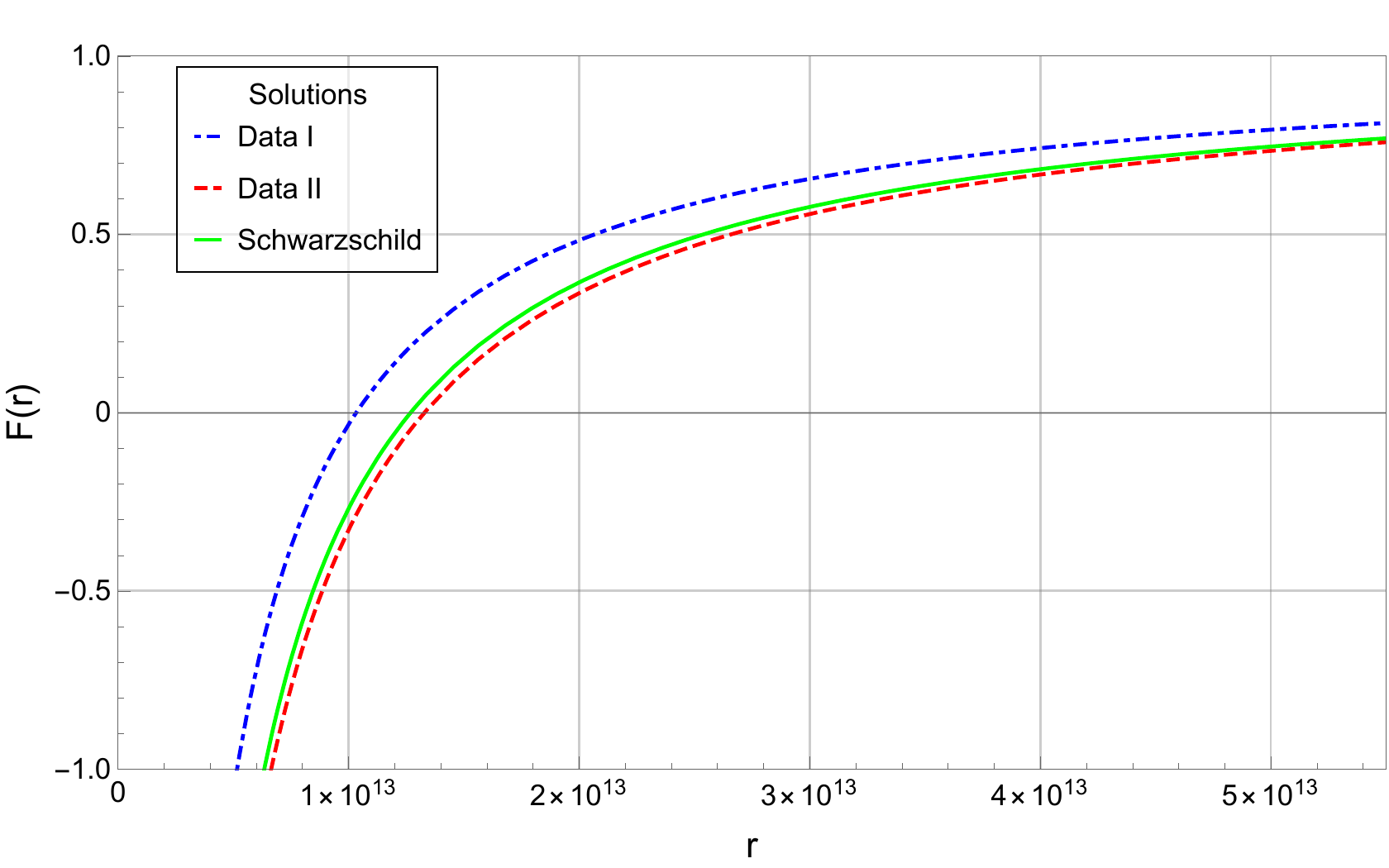}
\caption{Graphical representation of $F(r)$ from Eq.~(\ref{eq_f_sol_03}). For parameter sets: ($V_{cI}, a_I, M_I$) given by Data I, ($V_{cII}, a_{II}, M_{II}$) given by Data II, and for the Schwarzschild-type solution without a halo (0,0,$M_0$).} 
\label{fig:F(r)_vs_r_model03}
\end{figure}

Figure~\ref{fig:F(r)_vs_r_model03} shows the behavior of $F(r)$, related to the temporal component of Eq.~(\ref{ds_sol03}), and shows slight changes in the values of $r_h$ for each data set. For Data I, Data II and without halo, respectively: $r_{hI}=1.034\times10^{13}\textnormal{m}$, $r_{hII}=1.33\times10^{13}\textnormal{m}$ and $r_{h}^{\textnormal{(sch)}}=1.27\times10^{13}\textnormal{m}$. The radii $r_{hI}$ and $r_{hII}$ were computed using Eq.~(\ref{rh3}), based on the parameters from Data I and Data II, as specified above Table~\ref{tab:parametros}. It is important to note that, in the case without a halo, corresponding to the standard Schwarzschild solution, we employed the mass value $M_0$, defined just below Table~\ref{tab:parametros}. The presence of the dark matter (DM) halo influences the value of $r_h$, indicating that higher values of the parameters $(m_{\rm BH}, V_c, a)$ lead to an increase in the radius compared to the halo-free case. Conversely, for lower values of these parameters, the radius decreases. The variation in the horizon radius can be related to changes in mass. Specifically, the inequalities $M_{I} < M_0 < M_{II}$ and $r_{hI} < r_{h}^{\text{(sch)}} < r_{hII}$ indicate that, for a spherical object with constant matter density, an increase in mass corresponds to an increase in its total surface area. This observation leads us to consider the influence of the halo on the black hole entropy $S$, which will be addressed in Subsection~\ref{subsec-Thermodynamic}.

\subsection{Kretschmann Scalar}

The Kretschmann scalar $K(r)$ for this model, taking into account Eq.~(\ref{K_escalar_00}), is given by
\begin{widetext}
\begin{eqnarray}
    K(r)&=&\frac{4}{r^4}\left[1-g(r)^{-1}\right]^2
    +
   \frac{2  [\lambda +(\xi -1) r]^2}{r^4g(r)^{2}(r-\lambda
   )^2} +\frac{\left\{a^4
   \left[\tau_1(r)-\tau_4(r)\right]+2 a^2 r^2 \left[\tau_1(r)+\tau_2(r)\right]+r^4 \tau_1(r)\right\}^2}{8r^6\left(a^2+r^2\right)^4
   (r-\lambda )^4}
   \nonumber \\
   && \hspace{-1cm}
   +\frac{8 \left[m_{BH}\left(a^2+r^2\right)^2 - 2a^2r^3V_c^2 \right]^2}{r^6\left(a^2+r^2\right)^4} +\frac{1}{8r^6}\left(\frac{2 [\lambda +(\xi -1) r] \left[m_{BH}(a^2+r^2)^2 - 2a^2r^3V_c^2 \right]}{(r-\lambda )\left(a^2+r^2\right)^2}
   +\frac{r\tau_3(r)}{g(r)(r-\lambda )^2}\right)^2 ,
\label{K(r)_sol03}
\end{eqnarray}
\end{widetext}
where, for simplicity, we define the expression as follows
\begin{equation}
	\tau_1(r)=2 m_{BH} \left[4 \lambda ^2+\left(\xi ^2-5 \xi +4\right) r^2+\lambda 
	(3 \xi -8) r\right],
\end{equation}
\begin{eqnarray}
	\tau_2(r)/r &=& (\xi
	-1) r^2 \left[-\xi +(\xi -1) V_c^2+3\right]
		\nonumber \\
	&&-2 \lambda  r \left(\xi
	+V_c^2-3\right)+\lambda ^2 \left(V_c^2-3\right),
\end{eqnarray}
\begin{eqnarray}
	\tau_3(r)&=&[\lambda +(\xi -1) r]^2-2 r [\lambda +(\xi -1) r]
		\nonumber \\
	&& \hspace{-0.75cm} -2(r-\lambda ) [\lambda +(\xi -1) r]+2 (\xi -1) r (r-\lambda) ,
\end{eqnarray}
and 
\begin{equation}
	\tau_4(r)=r \left[3 \lambda ^2+\left(\xi ^2-4 \xi +3\right) r^2+2
	\lambda  (\xi -3) r\right].
\end{equation}

Expanding Eq.~(\ref{K(r)_sol03}) as a Taylor series, and considering the behavior of $K(r)$ for large values of $r$, we obtain the following expressions:
\begin{eqnarray}
K(r)=\frac{1}{r^4}\Bigg[\left(\xi-1\right)^2\left(\xi^2-6\xi+17\right)\left(\frac{1}{4}-V_c^2\right)
		\nonumber \\
	&&\hspace{-7cm}
    +V_c^4\left(\xi^4-8\xi^3+30\xi^2-40\xi+33\right)\Bigg]+\mathcal{O}\left(\frac{1}{r^5}\right) ,\label{exp_far_K}
\end{eqnarray} 
and when $r$ is very small, we have the following approximation:
\begin{equation}
    K(r)=\frac{48m_{BH}^2}{r^6}+\frac{20m_{BH}}{r^5}\left(\frac{2m_{BH}\xi}{\lambda}-1\right)+\mathcal{O}\left(\frac{1}{r^4}\right) \,. 
    \label{exp_near}
\end{equation}
For the Schwarzschild solution, we have the scalar $K_0(r)$,
\begin{equation}
    K_0(r)=\frac{48m^2_{BH}}{r^6}\label{K_schw} \ .
\end{equation}

\begin{figure}[htb!]
\includegraphics[scale=0.28]{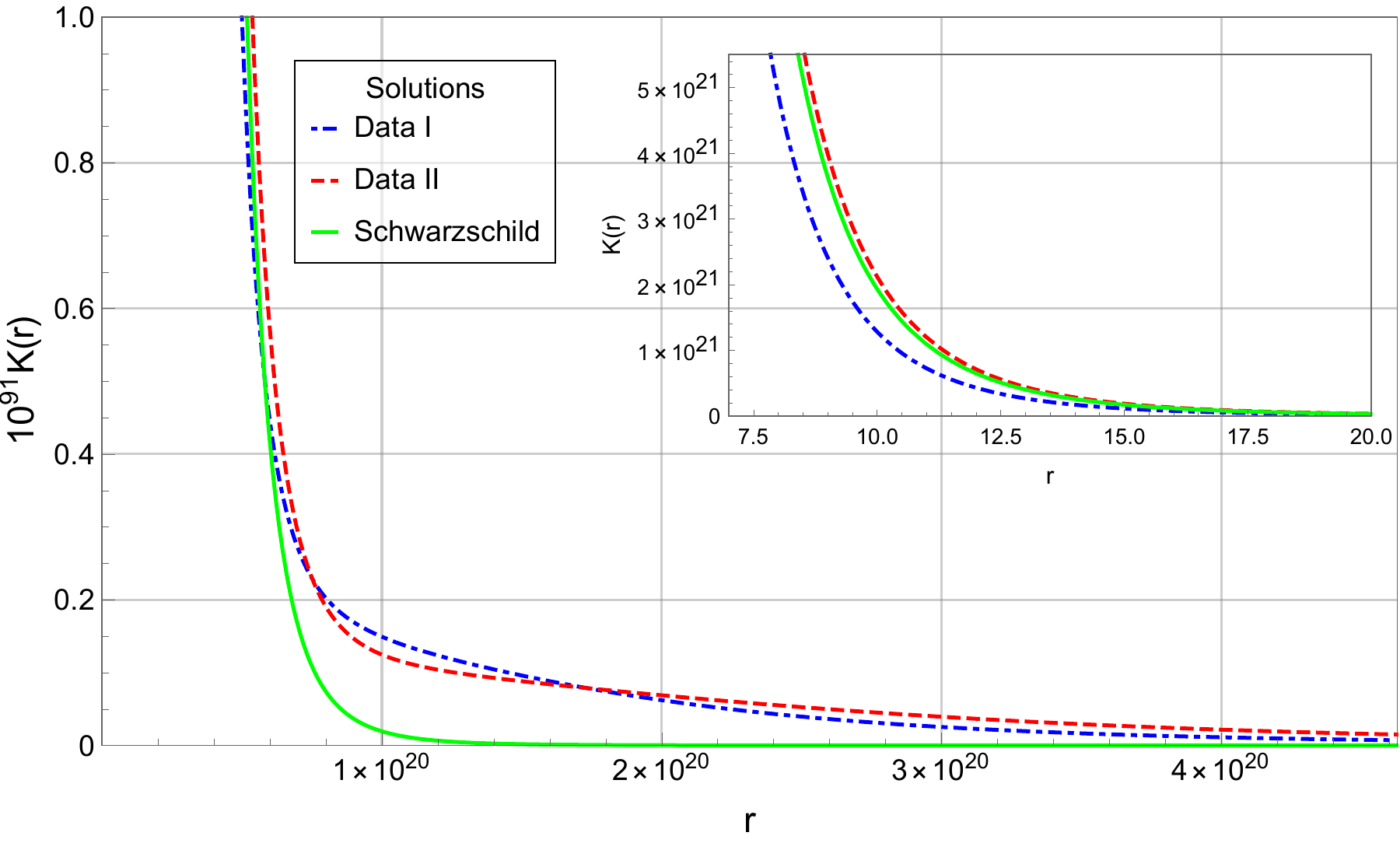}
\caption{Graphical representation of the scalar $K(r)$ from Eq.~(\ref{K(r)_sol03}). For the value sets: ($V_{cI},a_I,M_I$), given by Data I, ($V_{cII},a_{II},M_{II}$), given by Data II, and for the Schwarzschild-type solution without a halo (0,0,$M_0$). These values are listed immediately before and after Table \ref{tab:parametros}. The main graph shows the behaviour for larger scales of $r$, and the inner graph refers to smaller scales of $r$.} 
\label{fig:K(r)_vs_r_model03}
\end{figure}

For regions sufficiently far from the black hole center, Eq.~(\ref{exp_far_K}) shows that $K(r) \sim \frac{1}{r^{4}}$, indicating that the curvature decreases with increasing radius, as expected for asymptotically flat spacetimes. This implies that the halo’s contribution becomes negligible in this regime, vanishing regardless of the halo parameter values as $r \to \infty$. This behavior is confirmed by the main plot in Fig.~\ref{fig:K(r)_vs_r_model03}.

In contrast, for regions near the black hole center, Eq.~(\ref{exp_near}) and the inset in Fig.~\ref{fig:K(r)_vs_r_model03} show that the dominant term behaves as $\sim 1/r^6$, characteristic of the Schwarzschild solution. This indicates that the central singularity is not eliminated by the presence of the halo. Nonetheless, the halo contributes through subleading terms such as $\sim 1/r^5$ and others of lower order, primarily influenced by the parameters $V_c$ and $a$.

Figure~\ref{fig:K(r)_vs_r_model03} shows that $K(r)$, with a halo, retains the asymptotic behavior for Data I and II, analogous to the case without halo. From Eq.~(\ref{K(r)_sol03}) it generally follows that
\begin{equation}
\lim_{r \to \infty} K(r)\to0 \,, \qquad {\rm and} \qquad \lim_{r \to 0} K(r)\to \infty \,,
\label{limit_K}
\end{equation}
is a BH solution with a singularity in which spacetime becomes asymptotically flat at large distances from its center.

Analogous to the case without a halo, Fig.~\ref{fig:K(r)_vs_r_model03} shows that for higher values of the parameters $(m_{BH},V_c,a)$, which correspond to Data II, the values of $(K(r),r)$ are slightly larger. For Data I, on the other hand, the lower values of the same parameters indicate a slight decrease in curvature, indicated by a shift to the left. This suggests a relationship between mass and curvature, where BHs with larger mass, which is increased by the halo, exhibit curvature that extends over a larger surrounding area. The reverse behaviour is observed for smaller masses.

The overall behavior of $K(r)$ is not only illustrated in Fig.~\ref{fig:K(r)_vs_r_model03}, but also presented numerically in Table~\ref{tab:K_gradual}, which highlights the differential behavior at scales near the black hole center and the transition toward the characteristic radius $a$ of the halo. Around $r \sim a$, the curvature exhibits a slower decay in the presence of the halo compared to the halo-free case. This indicates that, while the datasets reflect distinct curvature behavior near the central region, they also reveal a pronounced increase in curvature extending into the vicinity of the halo’s critical radius. In essence, the halo prolongs the persistence of the spacetime curvature, although this effect vanishes asymptotically, as described by Eq.~(\ref{limit_K}).

\begin{table}[h]
    \centering
    \renewcommand{\arraystretch}{1.2}
    \begin{tabular}{lcc}
        \toprule
        \textbf{$r$[m]} & \textbf{$K_0(r)/K_I(r)$} & \textbf{$K_0(r)/K_{II}(r)$} \\
        \midrule
        $10^{-7}r_{h}^{\textnormal{(sch)}}=1.27\times10^6$ & $1.50857$ & $0.91181$ \\
        $r_{h}^{\textnormal{(sch)}}=1.27\times10^{13}$ & $1.50857$ & $0.911804$ \\
        $4\times10^6r_{h}^{\textnormal{(sch)}}=5.08\times10^{19}$ & $1.2208$ & $0.832969$ \\
        $0.19a_I=5.8634\times10^{19}$ & $0.980184$ & $0.747298$ \\
        $a_I=3.086\times10^{20}$ & $0.000955153$ & $0.000599851$ \\
        $a_{II}=4.629\times10^{20}$ & $0.000279914$ & $0.000132006$ \\
        $2a_{II}=9.258\times10^{20}$ & $0.0000484906$ & $0.0000179377$ \\
        \bottomrule
    \end{tabular}
\caption{Ratio between the scalar $K_0(r)$, given by Eq.~(\ref{K_schw}) for the Schwarzschild solution without a halo, and the scalar $K(r)$ corresponding to the halo-influenced solution, defined in Eq.~(\ref{K(r)_sol03}). The functions $K_I(r)$ and $K_{II}(r)$ are derived from Data~I and Data~II, respectively.}\label{tab:K_gradual}
\end{table}

Examining Table~\ref{tab:K_gradual}, particularly the first three rows corresponding to $r \sim r_{h}^{\text{(sch)}}$, we observe an increase in curvature intensity for Data~II, in contrast to the attenuation observed for Data~I. This behavior highlights the influence of the parameters $V_c$ and $a$ at scales on the order of $r_{h}^{\text{(sch)}}$, given that $(V_{cII}, a_{II}) > (V_{cI}, a_{I})$. Moreover, the inset of Fig.~\ref{fig:K(r)_vs_r_model03} shows that even near the black hole center, where the curvature diverges, the scalar $K(r)$ remains sensitive to the halo parameters. Accordingly, considering the dark matter density profile $\rho_{\rm DM}$ given in Eq.~(\ref{densidade}), such an influence is expected to manifest prominently within the halo's central core region.

\begin{figure}[htb!]
\includegraphics[scale=0.33]{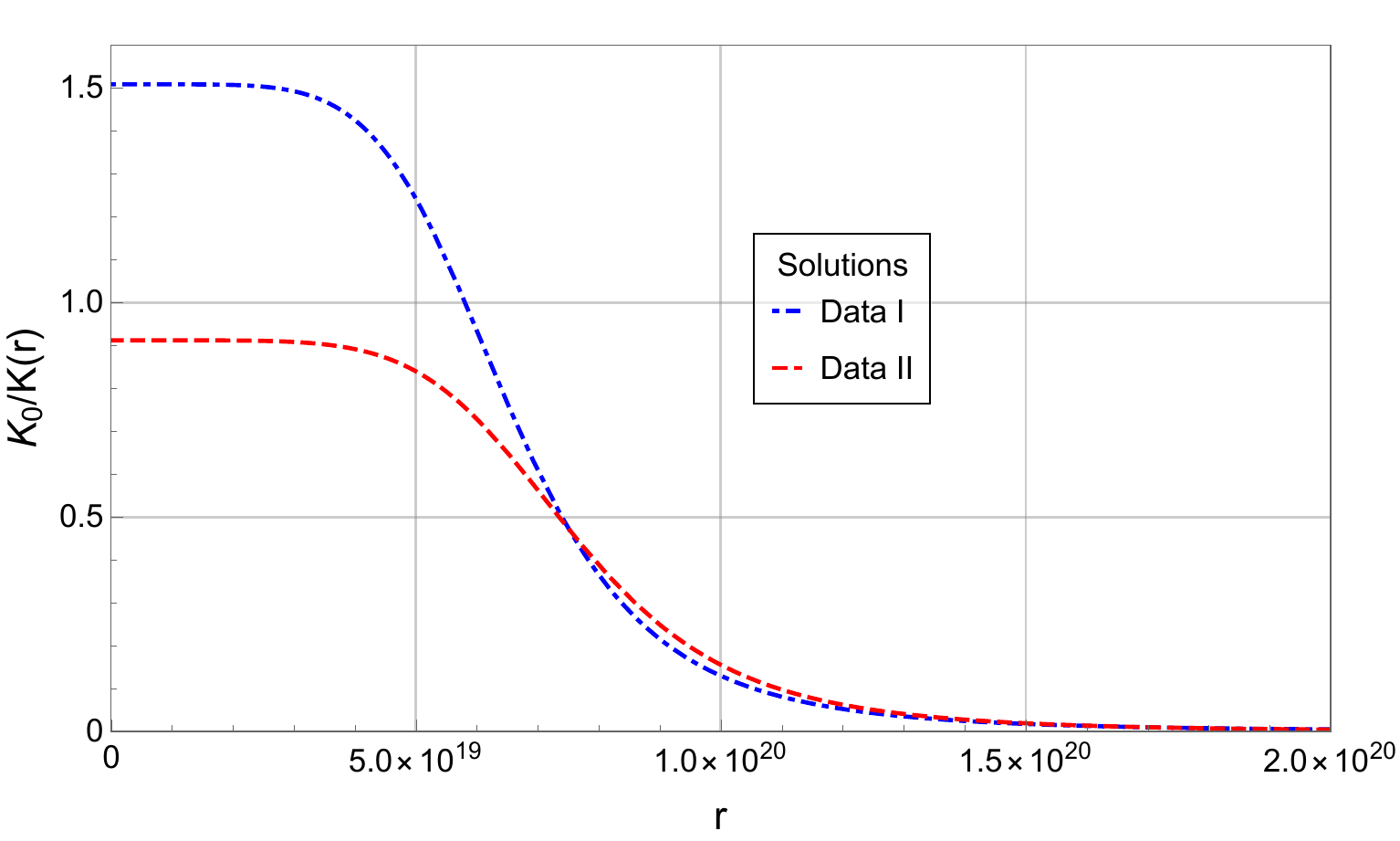}
\caption{Graphical representation of the ratio $K_0(r)/K(r)$, where $K(r)$ corresponds to the solution with a DM halo as given in Eq.~(\ref{K(r)_sol03}), and $K_0(r)$ represents the Schwarzschild solution without a halo from Eq.~(\ref{K_schw}). The plots are generated using the parameter sets $(V_{cI}, a_I, M_I)$ and $(V_{cII}, a_{II}, M_{II})$, corresponding to Data~I and Data~II, respectively. These values are detailed immediately before and after Table~\ref{tab:parametros}.} 
\label{fig:K(r)_ratio}
\end{figure}

Figure~\ref{fig:K(r)_ratio} presents the ratio $K_0(r)/K(r)$, previously discussed in Table~\ref{tab:K_gradual}, illustrating how the parameters $V_c$ and $a$ contribute to an increase in curvature as one approaches the black hole center. This ratio stabilizes within the halo’s ``core'' region, where it reaches the constant value $K_0(r)/K(r) = M_0/m_{\rm BH}$. As shown in Table~\ref{tab:K_gradual}, we find $K_0(r) = 1.50857\,K_I(r)$ for Data~I and $K_0(r) = 0.91181\,K_{II}(r)$ for Data~II. These results indicate that larger values of the halo parameters—i.e., $(V_{cII}, a_{II}) > (V_{cI}, a_{I})$—lead to a reduction in the curvature within the halo’s core.

\subsection{Black Hole Shadow Radius}

A distinctive observational feature of BHs is their \textit{shadow}, which arises due to the bending of light near the photon sphere, which is a spherical region where massless particles can orbit the BH in unstable circular trajectories. The presence of such a region allows a distant observer, located at radial coordinate $r_O$, to detect the BH's shadow \cite{Vagnozzi:2022moj}.

To determine the shadow radius, we adopt the formalism presented in \cite{Vagnozzi:2022moj, Perlick:2021aok}, starting with the spacetime metric introduced in Eq.~(\ref{metrica}). We define the auxiliary function $h(r)$ as
\begin{equation}
	h(r) = \sqrt{\frac{r^2}{f(r)}} \,,
\end{equation}
which characterizes the impact parameter of photons.

The radius of the photon sphere, denoted $r_{ph}$, is obtained by identifying the extremum of $h^2(r)$, i.e.,
\begin{equation}
	\left. \frac{d h^2(r)}{dr} \right|_{r = r_{ph}} = 0 \,.
\end{equation}
This condition yields the following differential relation involving the metric coefficient $F(r)$:
\begin{equation}
	\left. \left[F(r) - \frac{r}{2} \frac{dF(r)}{dr} \right] \right|_{r = r_{ph}} = 0 \,, 
	\label{r_ph__metodo}
\end{equation}
which determines $r_{ph}$ in terms of the geometry.

Once the photon sphere is identified, the shadow radius $r_{sh}$ observed at a radial coordinate $r_O$ is given by
\begin{equation}
	r_{sh} = r_{ph} \sqrt{ \frac{F(r_O)}{F(r_{ph})} } \,. \label{r_sh_GERAL_naoplana}
\end{equation}
In the limit where the spacetime becomes asymptotically flat, i.e., $F(r_O) \to 1$ as $r_O \to \infty$, this expression simplifies to
\begin{equation}
	r_{sh} = \frac{r_{ph}}{ \sqrt{F(r_{ph})} } \,. 
	\label{r_sh_GERAL}
\end{equation}

It is important to note that both the photon sphere radius and the corresponding shadow radius must lie outside the event horizon to be observable \cite{Vagnozzi:2022moj}.

By substituting the temporal metric coefficient given in Eq.~(\ref{ds_sol03}) into the condition for the photon sphere, Eq.~(\ref{r_ph__metodo}), we obtain the photon sphere radius $r_{ph}$ as
\begin{equation}
	r_{ph} = \frac{3\lambda}{3 - \xi} \,.
\end{equation}
Using this expression in Eq.~(\ref{r_sh_GERAL_naoplana}), the corresponding black hole shadow radius $r_{sh}$ is then given by
\begin{equation}
	r_{sh} = 3\sqrt{3} \sqrt{ \frac{\lambda^3}{(3 - \xi)^3} \left( \frac{\lambda \xi}{3 - \xi} \right)^{-\xi} F(r_O) } \,, \label{Sombra_sol03_FO}
\end{equation}
where $F(r_O)$ denotes the value of the metric function at the observer's location $r_O$.

\begin{figure}[htb!]
\includegraphics[scale=0.28]{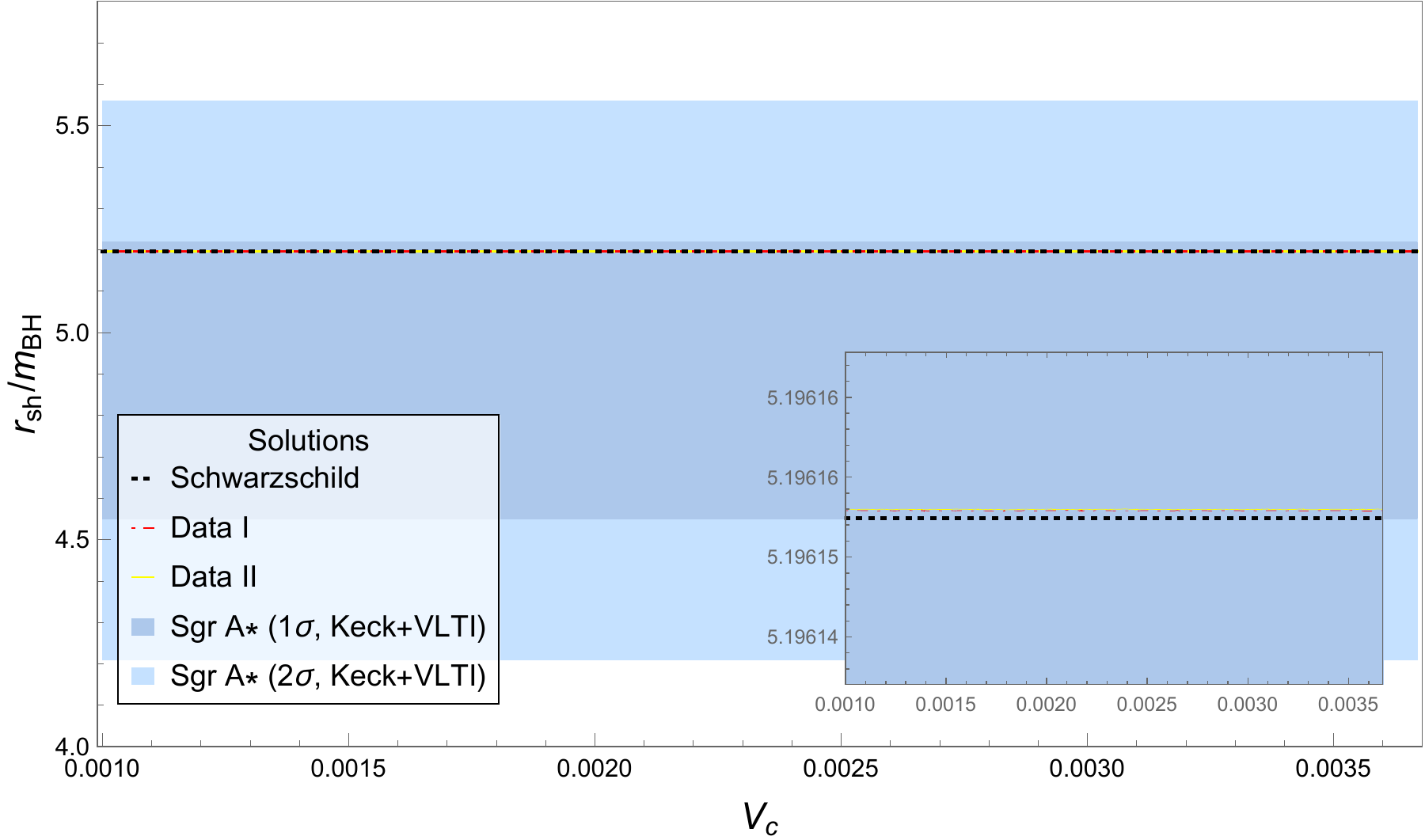}
\caption{Graphical representation of the ratio $r_{sh}/m_{BH}$, from Eq.~(\ref{Sombra_sol03_FO}), for $V_c$ values within the $1\sigma$ and $2\sigma$ measurement intervals. For the parameter sets: ($a_I,M_I$) given by Data I, ($a_{II},M_{II}$) given by Data II and (0,$M_0$) for the Schwarzschild-type solution without halo.} 
\label{fig:rshadow_vs_velocidade_model03}
\end{figure}

Figure~\ref{fig:rshadow_vs_velocidade_model03} illustrates the variation of $r_{sh}$ with respect to the critical velocity $V_c$, as defined in Table~\ref{tab:parametros}. The model $F(r)$ remains within the expected bounds for the specified range of $V_c$. It is observed that the presence of the halo leads to a slight increase in the shadow radius, independent of the specific value of $V_c$. This behavior can be attributed to the additional gravitational influence exerted by the halo, effectively increasing the gravitational mass and causing photons to bend at larger distances from the BH.

\begin{figure}[htb!]
\includegraphics[scale=0.28]{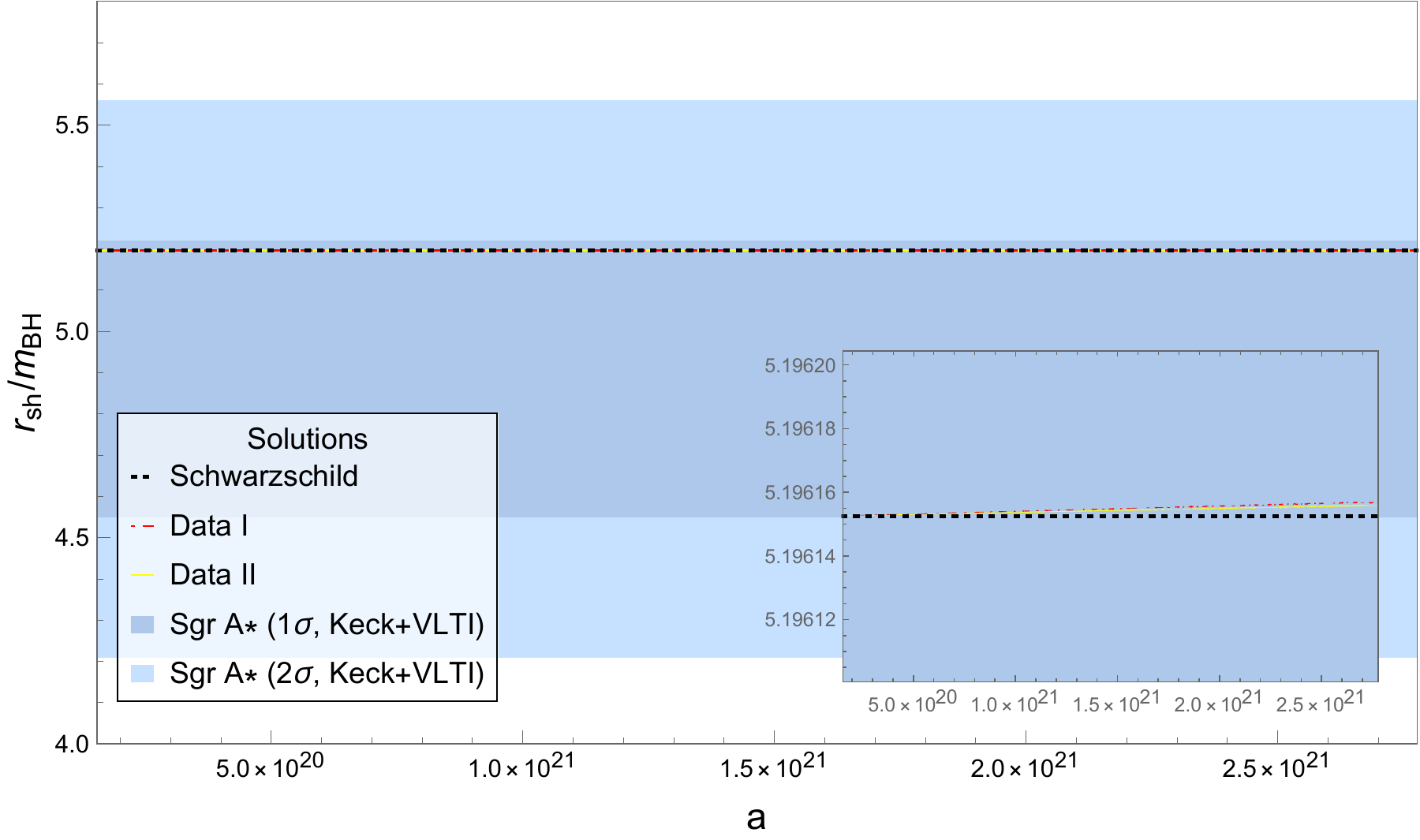}
\caption{Graphical representation of the ratio $r_{sh}/m_{BH}$, from Eq.~(\ref{Sombra_sol03_FO}), for $a$ values within the $1\sigma$ and $2\sigma$ measurement intervals. For the parameter sets: ($V_{cI},M_I$), given by Data I, ($V_{cII},M_{II}$), given by Data II, and (0,$M_0$) for the Schwarzschild-type solution without halo.} 
\label{fig:rshadow_vs_raioa_model03}
\end{figure}

In Figure~\ref{fig:rshadow_vs_raioa_model03}, we present the dependence of the dimensionless ratio $r_{sh}/m_{\rm BH}$ on the halo radius $a$, across the range specified in Table~\ref{tab:parametros}. The plot indicates an approximately proportional relationship, $r_{sh} \propto a$, which reflects a subtle enhancement in the curvature near the BH due to the extended mass distribution. As the halo becomes more extended, the region of significant spacetime curvature broadens accordingly.

Numerically, this slight change can be recognized if we establish the following relationship:
\begin{equation}
\frac{r_{sh}(m_{BHI},V_{cI},a_I)/m_{BHI}}{r_{sh}^{(sch)}/M_0}-1=0.927\times10^{-7} \ ,\label{razaoI}
\end{equation}
and
\begin{equation}
\frac{r_{sh}(m_{BHII},V_{cII},a_{II})/m_{BHII}}{r_{sh}^{(sch)}/M_0}-1=1.059\times10^{-7} ,\label{razaoII}
\end{equation}
in each case for the quantities
$(M_{I},V_{cI},a_{I})$ and $(M_{II},V_{cII},a_{II})$. 

In Tables~\ref{tab:rsh_Vc} and~\ref{tab:rsh_a}, we apply the same approach as in expressions~(\ref{razaoI}) and~(\ref{razaoII}) to highlight the subtle modifications to the BH shadow induced by variations in $V_c$ and $a$ within the parameter ranges specified in Table~\ref{tab:parametros}. Specifically, Table~\ref{tab:rsh_Vc} examines the ratio $r_{sh}/m_{BH}$ under changes in $V_c$, while Table~\ref{tab:rsh_a} explores the same ratio under variations in $a$.

Table~\ref{tab:rsh_Vc} corresponds to the graphs shown in Fig.~\ref{fig:rshadow_vs_velocidade_model03}, demonstrating that although the shadow radius $r_{sh}$ varies with $V_c$, the values of ${\rm RatioI}(V_c)$ and ${\rm RatioII}(V_c)$ remain essentially constant. This indicates that within this interval, a non-zero $V_c$ is sufficient to produce an increase in $r_{sh}$. We note an apparent oscillation of ${\rm RatioI}(V_c)$ and ${\rm RatioII}(V_c)$ about stable averages $\overline{\rm RatioI}(V_c) = 0.932867 \times 10^{-7}$ and $\overline{\rm RatioII}(V_c) = 1.08344 \times 10^{-7}$, with corresponding standard deviations $\omega_I = 0.0187499 \times 10^{-7}$ and $\omega_{II} = 0.0115171 \times 10^{-7}$, respectively. Both standard deviations are small relative to their means, indicating low variability of the values reported in Table~\ref{tab:rsh_Vc}. Consequently, the coefficient of variation, given by $\frac{\omega_I}{\overline{\rm RatioI}(V_c)} \times 100\% \approx 2\%$ and $\frac{\omega_{II}}{\overline{\rm RatioII}(V_c)} \times 100\% \approx 1\%$, shows that ${\rm RatioII}(V_c)$ is more stable than ${\rm RatioI}(V_c)$. This observation aligns with \cite{Shen:2009my}, which suggests that the parameter values in Data~II are more precise than those in Data~I.

Table~\ref{tab:rsh_a} corresponds to the graphs in Fig.~\ref{fig:rshadow_vs_raioa_model03}, illustrating that both ratios increase with $a$, which suggests that an expansion of the halo’s extent leads to an increase in the shadow radius. At every value of $a$, the inequality ${\rm RatioI}(a) > {\rm RatioII}(a)$ holds, indicating that the Data~I configuration exhibits a greater variation in $r_{sh}$ compared to Data~II. This increased sensitivity to changes in the halo parameters aligns with the fact that $(V_{cII}, a_{II}) > (V_{cI}, a_{I})$, reflecting a stronger response of spacetime curvature. These findings are consistent with the behaviors observed in Figs.~\ref{fig:K(r)_vs_r_model03} and~\ref{fig:K(r)_ratio}.

\begin{table}[h]
    \centering
    \renewcommand{\arraystretch}{1.2}
    \begin{tabular}{lcc}
        \toprule
        \textbf{$V_c$} & \textbf{${\rm RatioI}(V_c)[\times10^{-7}]$} & \textbf{${\rm RatioII}(V_c)[\times10^{-7}]$} \\
        \midrule
        $0.001$ & $0.94926978944$ & $1.0848654197$ \\
        $0.0015$ & $0.91375943034$ & $1.0859227717$ \\
        $0.002$ & $0.90629899585$ & $1.0676928408$ \\
        $0.0025$ & $0.94801603900$ & $1.0958764474$ \\
        $0.003$ & $0.93327437556$ & $1.0719221444$ \\
        $0.0035$ & $0.94658315408$ & $1.0943817941$ \\
        
        \bottomrule
    \end{tabular}
    \caption{We define the functions ${\rm RatioI}(V_c) =\frac{r_{sh}(m_{BHI},V_{c},a_I)/m_{BHI}}{r_{sh}^{(sch)}/M_0}-1$ and ${\rm RatioII}(V_c) =\frac{r_{sh}(m_{BHII}, V_{c},a_{II})/m_{BHII}}{r_{sh}^{(sch)}/M_0}-1$ using $r_{sh}$ from Eq.~(\ref{Sombra_sol03_FO}), and $r_{sh}^{(sch)}/M_0=3\sqrt{3}$ from the Schwarzschild solution without halo. The values of $V_c$ are consistent with Table~\ref{tab:parametros}.}
    \label{tab:rsh_Vc}
\end{table}

\begin{table}[h]
    \centering
    \renewcommand{\arraystretch}{1.2}
    \begin{tabular}{lcc}
        \toprule
        \textbf{$a$[m]} & \textbf{${\rm RatioI}(a)[\times10^{-7}]$} & \textbf{${\rm RatioII}(a)[\times10^{-7}]$} \\
        \midrule
        $5\times10^{20}$ & $1.4972686757$ & $1.1580407233$ \\
        $1\times10^{21}$ & $3.1029277814$ & $2.4392042386$ \\
        $1.5\times10^{21}$ & $4.7085869159$ & $3.6875174048$ \\
        $2\times10^{21}$ & $6.2297376657$ & $4.7387284807$ \\
        $2.5\times10^{21}$ & $7.7508884422$ & $6.3155451291$ \\
        
        \bottomrule
    \end{tabular}
    \caption{We define the functions ${\rm RatioI}(a) =\frac{r_{sh}(m_{BHI},V_{cI},a)/m_{BHI}}{r_{sh}^{(sch)}/M_0}-1$ and ${\rm RatioII}(a) =\frac{r_{sh}(m_{BHII},V_{cII},a)/m_{BHII}}{r_{sh}^{(sch)}/M_0}-1$ using $r_{sh}$ from Eq.~(\ref{Sombra_sol03_FO}), and $r_{sh}^{(sch)}/M_0=3\sqrt{3}$ from the Schwarzschild solution without halo. The values of $a$ are consistent with Tabela~\ref{tab:parametros}.}
    \label{tab:rsh_a}
\end{table} 

Equations~(\ref{razaoI}) and~(\ref{razaoII}), together with Tables~\ref{tab:rsh_Vc} and~\ref{tab:rsh_a}, demonstrate that the difference in the shadow radius $r_{sh}$ between the halo solution and the Schwarzschild solution (without halo) is exceedingly small. Such a difference is experimentally challenging to detect given the precision limits of current observational instruments. In \cite{Hou:2018}, the authors studied the shadow of the Sgr A* black hole under DM density profiles different from the one considered here, specifically for the Cold Dark Matter (CDM) and Scalar Field Dark Matter (SFDM) models. Their results indicate that the presence of the halo causes a slight increase in the shadow radius, approximately $9.8 \times 10^{-3}\%$ for CDM and $9.8 \times 10^{-5}\%$ for SFDM. However, these effects are only significant for extreme values of the parameter $k \sim \rho_c R^3 \sim 10^7$, where $\rho_c$ is the central density and $R$ is the radius at which pressure and density vanish. In the present work, the calculated increases in $r_{sh}$ amount to approximately $0.926837 \times 10^{-5}\%$ for Data~I and $1.05949 \times 10^{-5}\%$ for Data~II. Therefore, although the presence of the dark matter halo does induce changes in the shadow radius, these variations are sufficiently subtle to be currently unobservable with existing measurement techniques.

\subsection{Thermodynamic Properties}\label{subsec-Thermodynamic}

The entropy $S$ of a BH is related to the area $A$ of its event horizon by the Bekenstein–Hawking formula \cite{Hawking:1975vcx}, given by $S=A/4$ where $A = 4\pi r_{h}^2$, so that
\begin{equation}
	S=\pi r^2_{h}\label{S}.
\end{equation}

Starting from the entropy relation as given in Eq.~(\ref{S}), we substitute $r_h$ into Eq.~(\ref{eq_f_sol_03}), resulting in
\begin{equation}
M(S,V_c,a)=\frac{\sqrt{S}\left(a^2\pi+S-2SV^2_c\right)}{2\sqrt{\pi}\left(a^2\pi+S\right)}\label{massa_entropia_sol03} \ .
\end{equation}

\begin{figure}[htb!]
\includegraphics[scale=0.297]{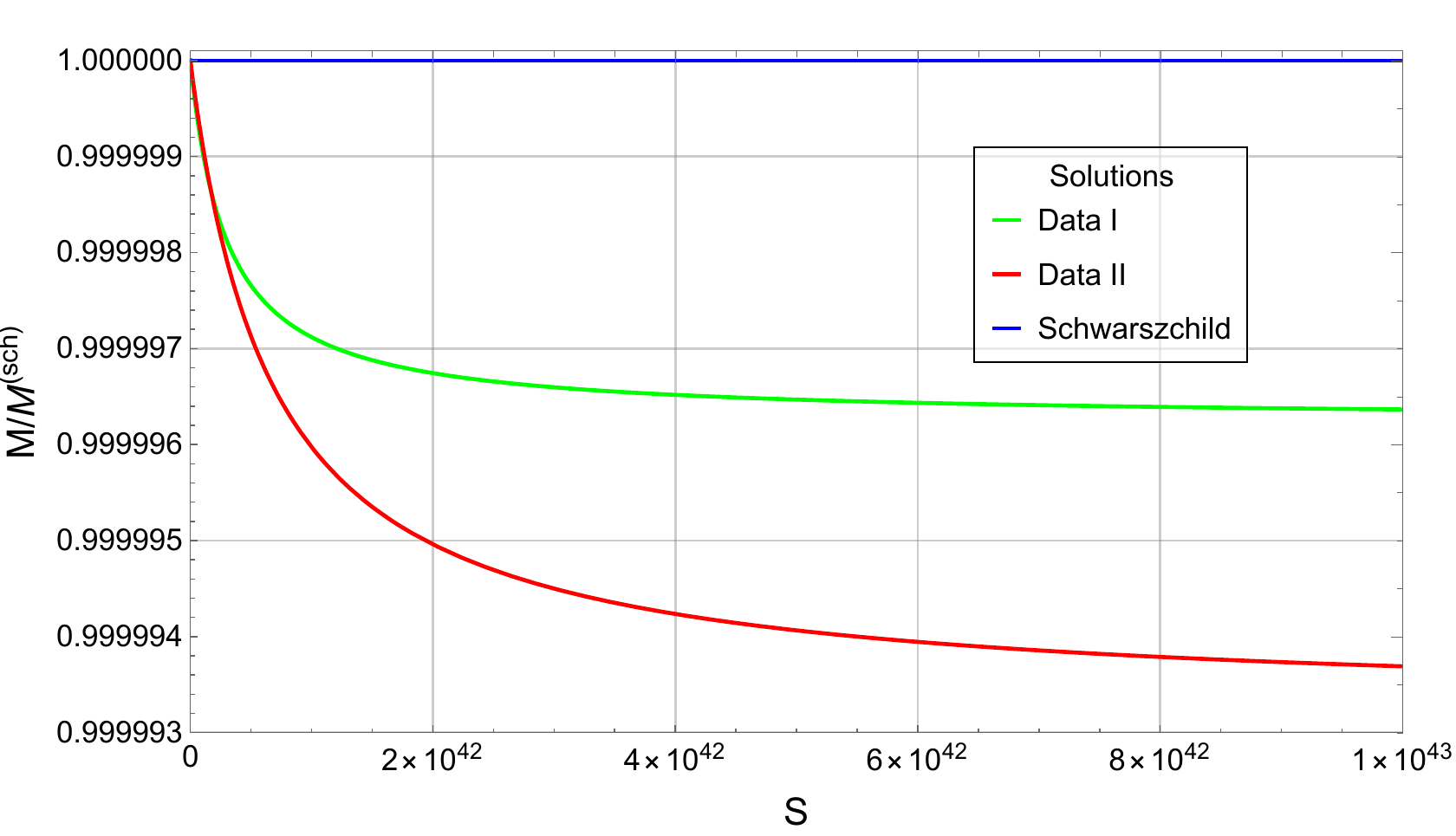}
\caption{Graphical representation of $M(S)/M^{(\textnormal{sch})}$, from Eq.~(\ref{massa_entropia_sol03}), where $M^{(\textnormal{sch})}$ refers to the solution of type SS. For the values: ($V_{cI},a_I$) given by Data I, ($V_{cII},a_{II}$) given by Data II and for the Schwarzschild-type solution without halo (0,0).} 
\label{fig:Massa_vs_S_model03}
\end{figure}

Figure~\ref{fig:Massa_vs_S_model03} illustrates the behavior of the mass ratio $M/M^{(\textnormal{sch})}$ as a function of the entropy $S$ for BH solutions embedded in a DM halo. Initially, this ratio remains close to unity, indicating that the presence of the halo does not significantly affect the mass. However, as $S$, $V_c$, and $a$ increase, the ratio decreases, leading to $M < M^{(\textnormal{sch})}$. This attenuation becomes more pronounced with larger values of the parameters, but eventually stabilizes, approaching a limiting value given by
\begin{equation}
	\lim_{S \to \infty} \frac{M}{M^{(\textnormal{Sch})}} = 1 - 2V_c^2 \,, \label{Mvs}
\end{equation}
which represents a lower bound for the mass reduction. Notably, Eq.~(\ref{Mvs}) reveals that the critical velocity $V_c$ of the DM halo is the dominant factor driving this effect.

To compute the temperature $T$ at the event horizon radius $r_h$, $T|_{r_h}=T$, we adopt the expression provided in \cite{Jiang:2007wj}, given by
\begin{equation}
	T = \left[ \frac{1}{4\pi} \sqrt{ \frac{dF(r)}{dr} \, \frac{d \left[1 / G(r) \right]}{dr} } \ \right]_{r = r_h} \,,
\end{equation}
where $F(r)$ and $G(r)$ are the temporal and radial metric functions, respectively. In this way we obtain
\begin{equation}
\scriptsize
T=\frac{\sqrt{\left[2a^2r_h^3V_c^2-m_{BH}(a^2+r_h^2)^2\right](r_h-\lambda)^{\xi-1}[\lambda+r_h(\xi-1)]}}{2\sqrt{2}\pi\left\{a^2(2m_{BH}-r_h)r_h+r_h^3[2m_{BH}+r_h(2V^2_c-1)]\right\}} \ .
\end{equation}
Since $\xi>1\to r_h=\lambda$, this leads to a disappearing temperature.
\begin{equation}
    T=0 \ .
\end{equation}

This corresponds to a cold BH solution or an extreme BH \cite{Preskill:1991tb,Ghosh:1995rv}. Therefore, the thermodynamic description for the proposed metric function seems to be inconsistent at first sight, as it tends to violate the third law of BH thermodynamics \cite{Bardeen:1973gs}. That is, according to $T=\kappa/2\pi$ \cite{Bardeen:1973gs}, a vanishing temperature implies a vanishing surface gravity, $\kappa = 0$. Since $S \propto r_h^2$, the entropy nevertheless remains non-zero. This peculiarity also occurs in the Kerr–Newman solution \cite{Newman:1965my,Adamo:2014baa} when the mass $M = \sqrt{\alpha^2 + Q^2}$, where $Q$ is the charge and $\alpha$ is a parameter associated with the angular momentum of the BH, and in the Kerr solution \cite{Alvarenga:2003jd} when $Q = 0$ and $M = \alpha$.

\subsection{Tangential Pressure}

\begin{figure}[htb!]
\includegraphics[scale=0.28]{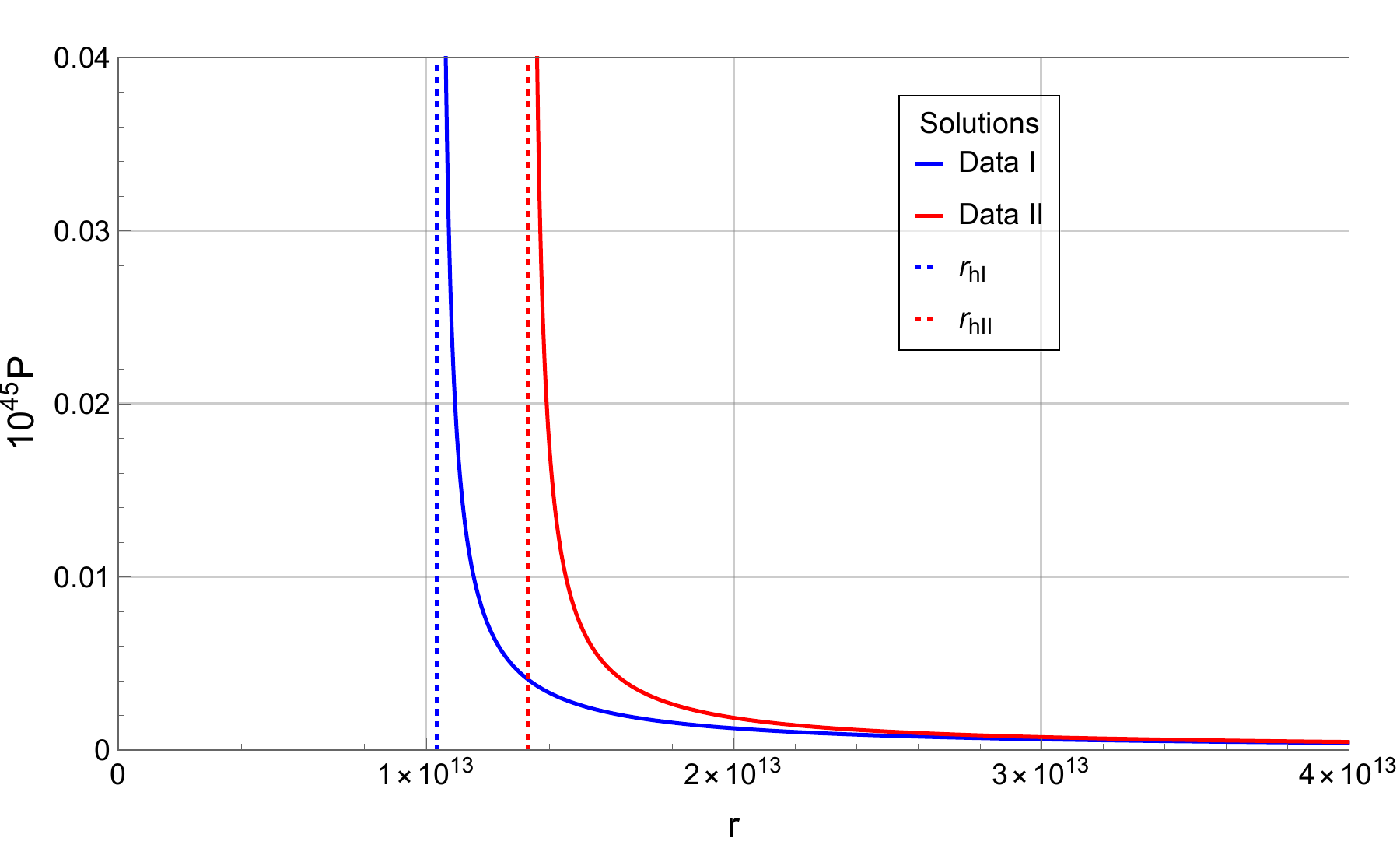}
\caption{Graphical representation of $P(r)$. For the values: ($M_{I},V_{cI},a_I$) given by Data I, and ($M_{II},V_{cII},a_{II}$) given by Data II.} 
\label{fig:pressao03}
\end{figure}

The tangential pressure $P(r)$, defined by Eq.~(\ref{pressao_03_DEF}), is explicitly given by
\begin{equation}
    P(r)=  \frac{V_c^2 \left(3 a^2+r^2\right) \left[m_{BH}(a^2 +r^2)+ r^3 V_c^2\right]}{8
   \pi  \left(a^2+r^2\right)^2 \left[\left(a^2+r^2\right) (r-2 m_{BH})-2 r^3
   V_c^2\right]} \ .\label{pressao_03}
\end{equation}
Figure~\ref{fig:pressao03} shows that the pressure becomes zero at large distances
\begin{equation}
\lim_{r\to\infty}P(r)\to0 \,,
\end{equation}
which is regular on the horizon, and assumes the following values: $P(r_{hI}=\lambda_I)=1.27\times10^{-41}\textnormal{m}^{-2}$ and $P(r_{hII}=\lambda_{II})=0.89\times10^{-41}\textnormal{m}^{-2}$ for Data I and II respectively. In IS units, we have $P(r_{hI}=\lambda_I)=1.5369\times10^{3}\ \text{kg}\ \text{m}^{-1}\ \text{s}^{-2}$ and $P(r_{hII}=\lambda_{II})=1.077\times10^{3}\ \text{kg}\ \text{m}^{-1}\ \text{s}^{-2}$. According to Eq.~(\ref{pressao_03_DEF}), if $r\to2M(r)$, then 
\begin{equation}
\lim_{r\to 2M(r)}P(r)\to\infty \ ,
\end{equation}
such that
\begin{equation}
    \left[r-M(r)\right]_{r=r_0}=0 \ ,
\end{equation}
where $r_0\lesssim r_h=\lambda$. The pressure thus characterizes a good behavior of this BH solution, since its divergence occurs in the region within $r_h$.
\par
For $r<r_h$ the signature and the components of the total energy-momentum tensor \cite{Wald:1984rg} change. Since the data in \cite{Shen:2009my} do not cover any aspects in the region within the event horizon, we do not analyze such a region in this paper.

\section{Conclusion}\label{sec:concl}

In this study, we have proposed and analyzed a Schwarzschild-type BH solution embedded within a DM halo, as characterized by the model introduced in \cite{Shen:2009my}. Our approach focuses on modeling the SMBH residing at the center of the galaxy NGC 4649 (also known as M60), under the influence of a surrounding DM distribution. By incorporating a DM halo profile into the spacetime geometry, we have investigated the resulting modifications to both the shadow characteristics and the thermodynamic properties of the BH. 

The considered DM halo profile is governed by two key parameters: the critical velocity $V_c$ and the critical radius $a$, which define the dynamics and spatial extent of the halo. This setup, inspired by earlier work in \cite{Cardoso:2021wlq}, introduces an additional contribution to the total mass function $M(r)$, rendering it radially dependent and sensitive to the properties of the halo. Importantly, this formulation departs from the standard Schwarzschild case by considering a non-vanishing $V_c$, thereby embedding the BH in a non-trivial astrophysical environment.

To explore the impact of these parameters on the BH properties, we adopted two observationally informed data sets—labeled Data I and Data II—derived from the parameter ranges established in \cite{Shen:2009my}. These data sets served as test beds to systematically evaluate how variations in $V_c$, $a$, and the central mass $M$ affect the spacetime structure, thermodynamic behavior, and observable features such as the shadow radius.

From a geometrical perspective, the solution features a single event horizon, the size of which is modulated by the DM halo via the influence of the three aforementioned parameters. We examined the Kretschmann scalar $K$, a curvature invariant used to diagnose the presence and nature of spacetime singularities. The analysis reveals a divergent behaviour as $r \to 0$, signalling the presence of a central singularity, and an asymptotic flatness as $r \to \infty$. In the intermediate regime, the curvature structure is significantly altered by the halo, indicating that the mass distribution induced by the halo strongly shapes the geometry of the surrounding spacetime. We evaluated the percentage variation in curvature as a function of radius using the ratio $K_0(r)/K(r)$, where we observed that in regions close to the center of the BH, curvature is influenced by the change in mass caused by the presence of the halo.

A key observational feature of BHs, namely their shadow, was also examined in this context. Building on methodologies established in \cite{Vagnozzi:2022moj}, we analysed how the radius of the BH shadow, $r_{sh}$, responds to variations in halo parameters. Our results show that $r_{sh}$ is particularly sensitive to changes in the halo scale radius $a$ and the critical velocity $V_c$, both of which affect the shadow. These findings are consistent with prior modelling in \cite{Shen:2009my}, and with the conclusions obtained by \cite{Hou:2018}. Therefore, although the radius of the shadow is barely affected by the presence of the halo, measuring it can provide valuable information about the DM environment surrounding astrophysical BHs.

In the thermodynamic domain, our analysis focused on how the presence of the DM halo modifies traditional relations, particularly the mass-entropy relation $M(S)$. Despite the additional mass contributions from the halo, the relation $M^2 \propto S$ is approximately preserved due to the low values of $V_c$ adopted in our model. Notably, the mass function shows a decreasing trend relative to the Schwarzschild case, such that for high entropy values, the BH mass asymptotically approaches a lower bound: $M \sim (1 - 2V_c^2)M^{(\text{sch})}$ as $S \to \infty$. This behaviour illustrates how the halo effectively alters the energy budget of the BH.

One striking feature of the model is the vanishing of the surface temperature, $T = 0$, which corresponds to a surface gravity $\kappa = 0$. This signals the emergence of a cold or extremal BH, which does not radiate thermally, a state of equilibrium induced by the presence of the DM halo. However, it is important to emphasize that such an extremal condition does not necessarily imply dynamical or thermodynamical stability, and warrants further investigation.

Additionally, we derived the radial dependence of the tangential pressure $P(r)$ associated with the DM halo. This pressure remains finite and well-behaved at the event horizon, ensuring physical regularity at this boundary. However, within the interior region bounded by the horizon ($r < r_h$), $P(r)$ diverges, suggesting that the core region experiences highly non-linear effects—possibly a limitation of the effective description or an indication of new physics required at small scales.

In summary, our results underscore the substantial impact that a surrounding DM halo can have on the physical, geometric, and thermodynamic properties of Schwarzschild-type BHs. These effects are not merely theoretical curiosities but carry potential implications for interpreting astrophysical observations, particularly those probing near-horizon phenomena such as BH shadows and quasi-periodic oscillations.

The work presented here opens several promising directions for future research. A critical next step involves assessing the stability of these DM-influenced BH solutions under perturbations—both linear and non-linear—to determine whether they represent physically realizable and long-lived configurations. Moreover, gravitational lensing by such BHs embedded in halos could offer additional, observable signatures of the DM influence, which may be detectable by next-generation telescopes. Another exciting avenue is the study of quasi-periodic oscillations in accretion disks, which could be used to infer the total mass and structure of such composite BH-halo systems. Through these future explorations, we aim to deepen our understanding of the interplay between dark matter and gravity in extreme environments, thereby contributing to the broader quest to reconcile GR with the unseen components of the cosmos.

\acknowledgments{MER thanks Conselho Nacional de Desenvolvimento
Cient\'{\i}fico e Tecnol\'ogico - CNPq, Brazil, for partial financial support. This study was financed in part by the
Coordena\c{c}\~{a}o de Aperfei\c{c}oamento de Pessoal de N\'{\i}vel Superior - Brasil (CAPES) - Finance Code 001. FSNL acknowledges support from the Funda\c{c}\~{a}o para a Ci\^{e}ncia
e a Tecnologia (FCT) Scientific Employment Stimulus contract with reference CEECINST/00032/2018, and
funding through the research grants UIDB/04434/2020,
UIDP/04434/2020 and PTDC/FIS-AST/0054/2021.}



\end{document}